\documentclass[preprint,floatfix] {revtex4} 
 
\usepackage{graphicx}
\usepackage{subfigure}
\begin{document}

\title{The generalized pseudospectral approach to the bound states of
Hulth\'en and Yukawa potentials}
\author{Amlan K. Roy}
\affiliation{Department of Chemistry, University of New Brunswick, 
Fredericton, NB, E3B 6E2, Canada}
\email{akroy@unb.ca}

\begin{abstract}
The generalized pseudospectral method is employed to calculate the bound states
of Hulth\'en and Yukawa potentials in quantum mechanics, with special emphases on 
higher excited states and stronger couplings. Accurate energy eigenvalues, 
expectation values and radial probability densities are obtained through a 
nonuniform and optimal spatial discretization of the radial Schr\"odinger equation. 
Results accurate up to thirteen to fourteen significant figures are reported 
for all the 55 eigenstates of both these potentials with $n\leq$10 
for arbitrary values of the screening parameters covering a wide range of 
interaction. Furthermore, excited states as high as up to $n=17$ have been 
computed with good accuracy for both these potentials. Excellent agreement 
with the available literature data has been observed in all cases. The $n>6$ 
states of Yukawa potential has been considerably improved over all other existing 
results currently available, while the same for Hulth\'en potential are reported 
here for the first time. Excepting the $1s$ and $2s$ states of Yukawa potential,
the present method surpasses in accuracy all other existing results in the stronger 
coupling region for all other states of both these systems. This offers a simple
and efficient scheme for the accurate calculation of these and other screened
Coulomb potentials.  
\end{abstract}
\maketitle

\section{Introduction}
The screened Coulomb potentials, 
\begin{equation}
V(r)=-\frac{Z}{r} \sum_{k=0}^{\infty} V_k (\lambda r)^k
\end{equation}
of which Hulth\'en and Yukawa potentials are two simple representatives, have
been of considerable interest in the context of many physical systems and a 
considerable amount of work has been devoted to study their numerous features
over the years. Z is identified as the atomic number when these are used in the 
context of atomic systems, while the screening parameter $\lambda$ has different
significance in different branches. The Hulth\'en potential [1] is one of the 
most important short-range potentials and has been used in the nuclear and 
particle physics [2-5], atomic physics [6-7], solid-state physics [8-9], chemical
physics [10], etc. This is also a special case of the Eckart potential. The Yukawa
potential [11], on the other hand, has found applications in approximating the 
effects of screening of nuclear charges by plasmas (commonly termed as the 
Debye-H\"uckel potential), shielding effect in the atoms and also in the 
solid-state physics (as the Thomas-Fermi potential), etc. 

These two potentials have several similarities; e.g., they are both Coulomb-like
for small $r$ and decay monotonically exponentially to zero for large $r$. 
Another distinctive feature of these potentials (in contrast to the Coulomb 
potentials) is the presence of limited number of bound states characterized by
the presence of the screening parameters; i.e., bound states exist only for
certain values of the screening parameter below a threshold limit (e. g., for the
Yukawa potential, this value has been accurately estimated as 1.19061227 
$\pm $0.00000004 [12] in atomic units). The former has the additional special 
property that it offers exact analytical solutions for $\ell=0$ states only, not 
for the higher partial waves [13]. Many formally attractive and efficient 
formalisms have been proposed for accurate determination of the eigenvalues, 
eigenfunctions as well as for the values of the critical screening parameters 
differing in complexity, accuracy and efficiency. The most notable of these are 
the variational calculations employing a multitude of basis functions [14-18,12], 
combined Pad\'e approximation and perturbation theory [19-21], shifted 1/N 
approximation along with many of its variants [22-27], dynamical-group approach 
[28], supersymmetric quantum mechanics [29], numerical calculations [30-31,18] 
and other works [32-33]. 

In the past few years, the generalized pseudospectral (GPS) method has been proved
to be a very powerful and efficient tool to deal with the static and dynamic 
processes of many-electron atomic/molecular systems characterized by the Coulomb 
singularities (see, for example, [34-37] and the references therein). Recently it 
has also been successfully applied to the power-law and logarithmic potentials 
[38], as well as the spiked harmonic oscillator with stronger singularity [39]. 
One of the objectives of this article is to extend and explore the regions of 
validity of this formalism to a different class of singularities, namely the 
screened Coulomb potentials thus covering a broader range of physical systems. In
an attempt to assess the performance and its applicability to such systems, we have
computed all the 55 eigenstates (1$\leq n \leq$10) of the Hulth\'en and Yukawa 
potentials and compared them with the available literature data wherever possible 
with an aim to study the spectra of these systems systematically. It may be noted
that although many accurate results are available for these potentials in the weaker
coupling region and for the lower states, there is a lack of good quality results in
the stronger region and for the higher states. In this work, we pay special attention
to both of these issues. Screening parameters of arbitrary field strengths (covering
both weak and strong limits of interaction) have been considered for given values of
$n$ and $\ell$ quantum numbers. To this end, accurate calculations have been 
performed on the eigenvalues, expectation values and radial probability densities of 
these two systems. As a further stringent test of the method, we calculate some very
high excited states (up to $n=17$) of these two systems which have been examined never
before. As will be evident in a later section, this method is indeed capable of 
producing excellent quality results comparable in accuracy to the other existing 
literature data for both these systems and in many cases (especially in the stronger
regions of coupling), indeed offers the best results. The article is organized as 
follows: Section II presents an outline of the theory and the method of calculation. 
Section III gives a discussion on the results while section IV makes some 
concluding remarks. 

\section{The GPS method for the solution of the Hulth\'en and Yukawa potentials}
\label{sec:method}
In this section, we present an overview of the GPS formalism within the nonrelativistic
framework for solving the radial Schr\"odinger equation (SE) of a single-particle 
Hamiltonian containing a Hulth\'en or Yukawa term in the potential. Only the 
essential steps are given and the relevant details may be found elsewhere ([34-39]
and the references therein). Unless otherwise mentioned, atomic units are employed
throughout this article. 

The radial SE can be written in the following form, 
\begin{equation}
\left[-\frac{1}{2} \ \frac{\mathrm{d^2}}{\mathrm{d}r^2} + \frac{\ell (\ell+1)} {2r^2}
+v(r) \right] \psi_{n,\ell}(r) = E_{n,\ell}\ \psi_{n,\ell}(r)
\end{equation}
where 
\begin{equation}
v(r)= -\frac{Z\delta e^{-\delta r}} {1-e^{-\delta r}} \ \ \ \ \mathrm{(Hulthen)}
\end{equation}
or
\begin{equation}
 v(r) = -\frac{Z e^{-\lambda r}}{r} \ \ \ \ \mathrm{(Yukawa)} 
\end{equation}
where $\delta$ and $\lambda$ denote the respective screening parameters whereas $n$ and
$\ell$ signify the usual radial and angular momentum quantum numbers respectively. Use
of a scaling transformation, $r \rightarrow r/Z$ gives the following well-known
relation, 
\begin{equation}
E(Z,\delta (\lambda)) = Z^2\ E(1, \delta (\lambda)/Z)
\end{equation}
Thus it suffices to study only the $Z=1$ case and this fact has been used in this 
work. The usual finite difference or finite element discretization schemes often 
require a large number of grid points to achieve good accuracy and convergence, 
often because of their uniform nature. The GPS formalism, in contrast, can give 
nonuniform and optimal spatial discretization with a significantly smaller number 
of grid points allowing a denser mesh at smaller $r$ and a coarser mesh at larger 
$r$ while maintaining similar accuracies in both regions. In addition, this is 
also computationally orders of magnitude faster. 

The first step is to approximate a function $f(x)$ defined in the interval 
$x \in [-1,1]$ by the N-th order polynomial $f_N(x)$ such that, 
\begin{equation}
f(x) \cong f_N(x) = \sum_{j=0}^{N} f(x_j)\ g_j(x),
\end{equation}
and this ensures that the approximation is \emph {exact} at the \emph {collocation 
points} $x_j$, i.e.,
\begin{equation}
f_N(x_j) = f(x_j).
\end{equation}
Here, we have used the Legendre pseudospectral method having $x_0=-1$, $x_N=1$, where
$x_j (j=1,\ldots,N-1)$ are obtained from the roots of the first derivative of the 
Legendre polynomial $P_N(x)$ with respect to $x$, i.e., 
\begin{equation}
P'_N(x_j) = 0.
\end{equation}
The $g_j(x)$ in Eq.~(6) are called the cardinal functions and given by,
\begin{equation}
g_j(x) = -\frac{1}{N(N+1)P_N(x_j)}\ \  \frac{(1-x^2)\ P'_N(x)}{x-x_j},
\end{equation}
These have the unique property, $g_j(x_{j'}) = \delta_{j'j}$. Now the semi-infinite
domain $r \in [0, \infty]$ can be mapped onto the finite domain $x \in [-1,1]$ by
the transformation $r=r(x)$. The following algebraic nonlinear mapping,
\begin{equation}
r=r(x)=L\ \ \frac{1+x}{1-x+\alpha},
\end{equation}
may be used, where L and $\alpha=2L/r_{max}$ are termed as the mapping parameters. 
Now, introduction of the following relation, 
\begin{equation}
\psi(r(x))=\sqrt{r'(x)} f(x)
\end{equation}
followed by a symmetrization procedure leads to the transformed Hamiltonian as below, 
\begin{equation}
\hat{H}(x)= -\frac{1}{2} \ \frac{1}{r'(x)}\ \frac{d^2}{dx^2} \ \frac{1}{r'(x)}
+ v(r(x))+v_m(x),
\end{equation}
where $v_m(x)$ is given by,
\begin{equation}
v_m(x)=\frac {3(r'')^2-2r'''r'}{8(r')^4}.
\end{equation}
The advantage is clear; this leads to a \emph {symmetric} matrix eigenvalue problem
which can be readily solved to give accurate eigenvalues and eigenfunctions by using
some standard available routines. Note that $v_m(x)=0$ for the particular 
transformation used here and finally one obtains the following set of coupled 
equations, 
\begin{widetext}
\begin{equation}
\sum_{j=0}^N \left[ -\frac{1}{2} D^{(2)}_{j'j} + \delta_{j'j} \ v(r(x_j))
+\delta_{j'j}\ v_m(r(x_j))\right] A_j = EA_{j'},\ \ \ \ j=1,\ldots,N-1,
\end{equation}
\end{widetext}
where
\begin{equation}
A_j  = \left[ r'(x_j)\right]^{1/2} \psi(r(x_j))\ \left[ P_N(x_j)\right]^{-1}.
\end{equation}
and the symmetrized second derivatives of the cardinal function, $D^{(2)}_{j'j}$
are given by,
\begin{equation}
D^{(2)}_{j'j} =  \left[r'(x_{j'}) \right]^{-1} d^{(2)}_{j'j} 
\left[r'(x_j)\right]^{-1}, 
\end{equation}
with
\begin{eqnarray}
d^{(2)}_{j',j} & = & \frac{1}{r'(x)} \ \frac{(N+1)(N+2)} {6(1-x_j)^2} \ 
\frac{1}{r'(x)}, \ \ \ j=j', \nonumber \\
 & & \nonumber \\
& = & \frac{1}{r'(x_{j'})} \ \ \frac{1}{(x_j-x_{j'})^2} \ \frac{1}{r'(x_j)}, 
\ \ \ j\neq j'.
\end{eqnarray}
A large number of tests have been performed to check the accuracy and reliability
of the method so as to produce ``stable'' results with respect to the variation of
the mapping parameters. In this way, a consistent set of parameter sets were chosen.
For the problems at hand $\alpha=25$ and $N=200$ seemed appropriate for all the 
states considered in this work while $R$ values were varied as required (see later). 
The results are reported only up to the precision that maintained stability. It may
be noted that all our results are {\em truncated} rather than {\em rounded-off}. 

\section{Results and Discussion}
Let us first examine the convergence of the calculated energy eigenvalues. As an 
illustration, consider the following two cases of $s$ states of the Hulth\'en 
potential which offer exact analytical results: (a) the ground state with 
$\delta=1.97$ (high screening) and (b) $12s$ (moderately high state) with 
$\delta=0.005$ (intermediate screening). Variation of the eigenvalues was monitored 
with respect to the radial distance $R$ keeping the other two parameters $\alpha$ 
and N fixed at 25 and 200 respectively. It was seen that $R=200$ or 300 a. u., was 
not suitable for either of these situations (although these may be sufficient 
for weaker screenings) and with $R=500$ a. u., a reasonable convergence can be 
achieved. However, a better convergence in both these cases requires at least $R=800$
a. u. And after that, the calculated results are stable with respect to $R$. As 
expected, for even higher states one would require a larger $R$; e. g., the $17s$
state with same $\delta=0.005$ requires $R=5000$ a. u. However, the results are
apparently less sensitive with respect to $N$, the total number of grid points; only
200 points is sufficient for all the calculations reported in this work. This is in
sharp contrast with the finite difference (FD) or finite element (FE) methods where 
one usually requires a substantially larger number of radial grid points to achieve 
good convergence for such singular systems. This is more so, if one uses a uniform 
discretization scheme. The present method does not suffer from such an uncomfortable
feature, for it offers equally accurate eigenfunctions both at small and the large
distances with significantly smaller grid points. For a given screening parameter,
within a particular $n$, the required $R$ increases with increasing $\ell$. For a
given state, larger screening parameter requires larger $R$. For example, the $9g$
states of Hulth\'en potential reach convergence with $R=500$ a. u., for 
$\delta=0.001, 0.005, 0.01$; but $\delta=0.02$ needed $R=1500$ a.u. Similar 
considerations hold equally good for the Yukawa potential. 

\begingroup
\squeezetable
\begin{table}
\caption {\label{tab:table1}Calculated negative eigenvalues E (in a.u.) of some
selected $s$ states of the Hulth\'en potential for different $\delta$ along with the
literature data. An asterisk denotes the exact analytical value, Eq. (18). Numbers 
in the parentheses denote $\delta_c$ values [18].} 
\begin{ruledtabular}
\begin{tabular}{ccllcll}
State & $\delta$ & \multicolumn{2}{c}{$-$Energy} & $\delta$ 
& \multicolumn{2}{c}{$-$Energy} \\ 
\cline{3-4} \cline{6-7}
   &  & This work   & Reference &  &  This work & Reference \\   \hline
$1s$(2.0) & 0.002 & 0.49900050000000 & 0.4990005\footnotemark[1],0.4990005*  
      & 1.97  & 0.00011249999999 & 0.0001125*                            \\
$2s$(0.5)  & 0.025 & 0.11281249999999 & 0.1128124999960\footnotemark[2],0.1128125*
      & 0.492 & 0.00003200000000 & 0.000032*                         \\
$3s$(0.222) & 0.002   & 0.05456005555555 & 0.05456006\footnotemark[1], 
0.054560055*$\cdots$  & 0.21  & 0.00016805555555 & 0.000168055*$\cdots$    \\ 
$16s$ (0.008) & 0.001 & 0.0014851250000  & 0.001485125*
      & 0.005 & 0.0002531250000  & 0.000253125*                    \\
$17s$ (0.007) & 0.001 & 0.0012662288062  & 0.0012662288062*
      & 0.005 & 0.0001332288062  & 0.0001332288062*                    \\
\end{tabular}
\end{ruledtabular}
\footnotetext[1] {Ref. [23].}
\footnotetext[2] {Ref. [31].}
\end{table}

Now let us consider the $\ell=0$ states of Hulth\'en potential which offer
exact analytical results [1,13] given by,
\begin{equation}
E_n^{\mathrm{exact}}=-\frac{\delta^2}{8n^2}\left[\frac{2}{\delta}-n^2\right]
\end{equation}
with $n^2 <2/\delta$. Table I presents the calculated eigenvalues for some 
representative $\ell=0$ states with $n=1-17$ at selected values of $\delta$. For each
of these states, two screening parameters are chosen; weak in the left- and strong in 
the right-hand side. The critical screening parameters ($\delta_c$) for S states, 
given by the simple relation, $\delta_c=2/n^2$ [18], are presented in the parentheses
in column 1 along with the states. The exact eigenvalues calculated from the above
are given with an asterisk at the end. As noted, in the weaker region as well as for
lower states, other theoretical results are available, while no reference theoretical 
results could be found for higher states as well as for the stronger coupling cases. 
Because of their exactly solvable nature, some of these states (especially the lower 
ones like $1s$ and $2s$) have received extra attention from various workers employing 
a variety of methods and we have quoted a few of them. Some of these include the Lie 
algebraic method [40], Pad\'e approximation [19], path integral formulation [41], 
shifted 1/N expansion [23], dynamical group theoretical consideration [28], 
generalized variational calculation [16] as well as the accurate numerical 
calculation [31], etc. First, we note that the calculated values completely coincide 
with the exact analytical results for all these states encompassing the whole range 
of interaction nicely. This amply
demonstrates the accuracy, reliability and potential of the methodology. For $n=1-5$,
shifted 1/N expansion results [23] are available in the weak coupling region and the 
present results are considerably superior to these values in all cases. For $n=1-2$, 
accurate numerical eigenvalues [31] are available for $\delta \leq 0.3$. Their $n=1$
results are significantly better than the $n=2$. For $1s$ states, results of [31] are
comparable to ours, but for the $2s$ states, current values are superior to [31]. Quite
accurate results have been reported [16] for $2s-6s$ states that improved the previous
results in the literature significantly by employing trial wave functions which were
linear combinations of independent functions. Results of [16] are better than ours in
the weak-coupling region, but in the stronger limit, our results are noticeably better
than [16]. It may be mentioned that we have enlarged the coupling region from all other
previous works and it is clear that the current results are so far the most accurate 
values in regions closest to the critical limit. For states with $n>6$, no other 
theoretical results are available in the literature so far and we hope that these 
results may be helpful in future studies. It may be mentioned that the energies of 
$n=15-17$ states are slightly less accurate than the other lower $s$ state energies.  

\endgroup
\begingroup
\squeezetable
\begin{table}
\caption {\label{tab:table2}Calculated negative eigenvalues E (a.u.) of the 
Hulth\'en potential for selected $\ell \neq 0, n=2-6$ states for various $\delta$
values along with the literature data. Numbers in the parentheses denote the 
$\delta_c$ values taken from [18].} 
\begin{ruledtabular}
\begin{tabular}{ccllccll}
State & $\delta$ & \multicolumn{2}{c}{$-$Energy} & State &  $\delta$ &  
\multicolumn{2}{c}{$-$Energy} \\ 
\cline{3-4} \cline{7-8}
   &  & This work   & Literature &  &   &  This work & Literature \\   \hline
$2p(0.377)$ & 0.005 & 0.12251041674479 &  &     $4f(0.086)$ 
            & 0.01  & 0.02640009031711 & 0.02640\footnotemark[2]  \\
            & 0.35  & 0.00379309814702 & 0.00379309814702\footnotemark[1] &   
            & 0.08  & 0.00135376897143 &                          \\      
            & 0.36  & 0.00220960766773 &  &     $5f(0.060)$ 
            & 0.005 & 0.01756564260992 &                          \\
$3p(0.186)$ & 0.005 & 0.05308159769106 &  &  
            & 0.05  & 0.00178354579471 & 0.00178354579471\footnotemark[1]  \\
            & 0.15  & 0.00446630878535 & 0.00446630878535\footnotemark[1] 
            & $5g(0.055)$ & 0.005 & 0.01755731319688              \\
            & 0.18  & 0.00047689388317 &      &   
            & 0.05  & 0.00101588159045 & 0.00101588159045\footnotemark[1]  \\
$3d(0.158)$ & 0.005 & 0.05307743154020 &      & $6g(0.041)$ 
            & 0.005 & 0.01148061249746 &                          \\
            & 0.15  & 0.00139659246573 & 0.00139659246573\footnotemark[1] &
            & 0.025 & 0.00372009346428 & 0.00372009346428\footnotemark[1]  \\ 
$4d$(0.098) & 0.075 & 0.00383453307692 & 0.00383453307692\footnotemark[1] & 
            $6h(0.038)$ & 0.005 & 0.01147020315553                         \\
            & 0.09  & 0.00099103405815 &      &   
            & 0.025 & 0.00346543458707 & 0.00346543458707\footnotemark[1]  \\
\end{tabular}
\end{ruledtabular}
\footnotetext[1] {Ref. [16].}
\footnotetext[2] {Ref. [33].}
\end{table}
\endgroup

\begin{figure}
\begin{minipage}[c]{0.40\textwidth}
\centering
\includegraphics[scale=0.45]{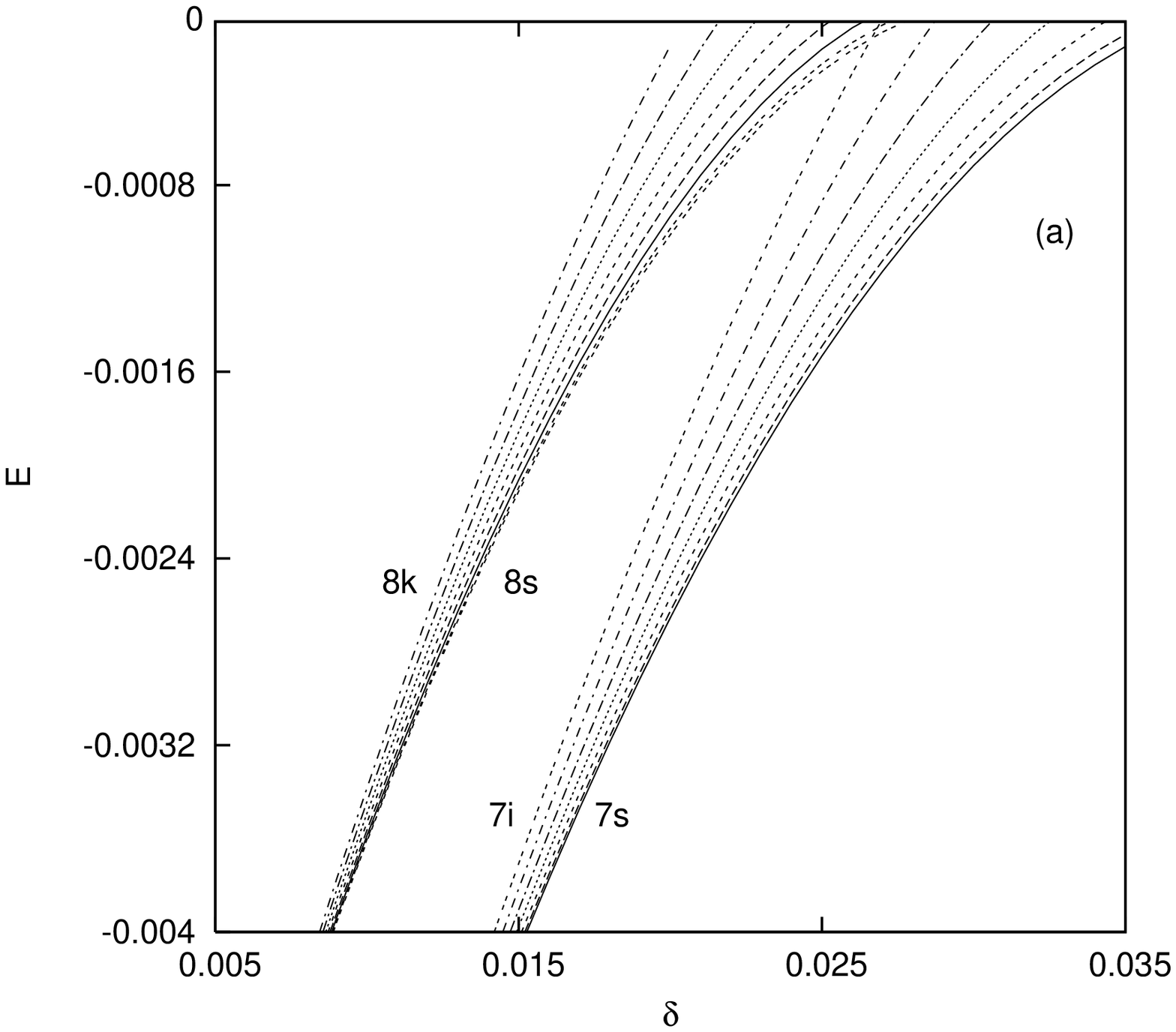}
\end{minipage}%
\hspace{0.5in}
\begin{minipage}[c]{0.40\textwidth}
\centering
\includegraphics[scale=0.45]{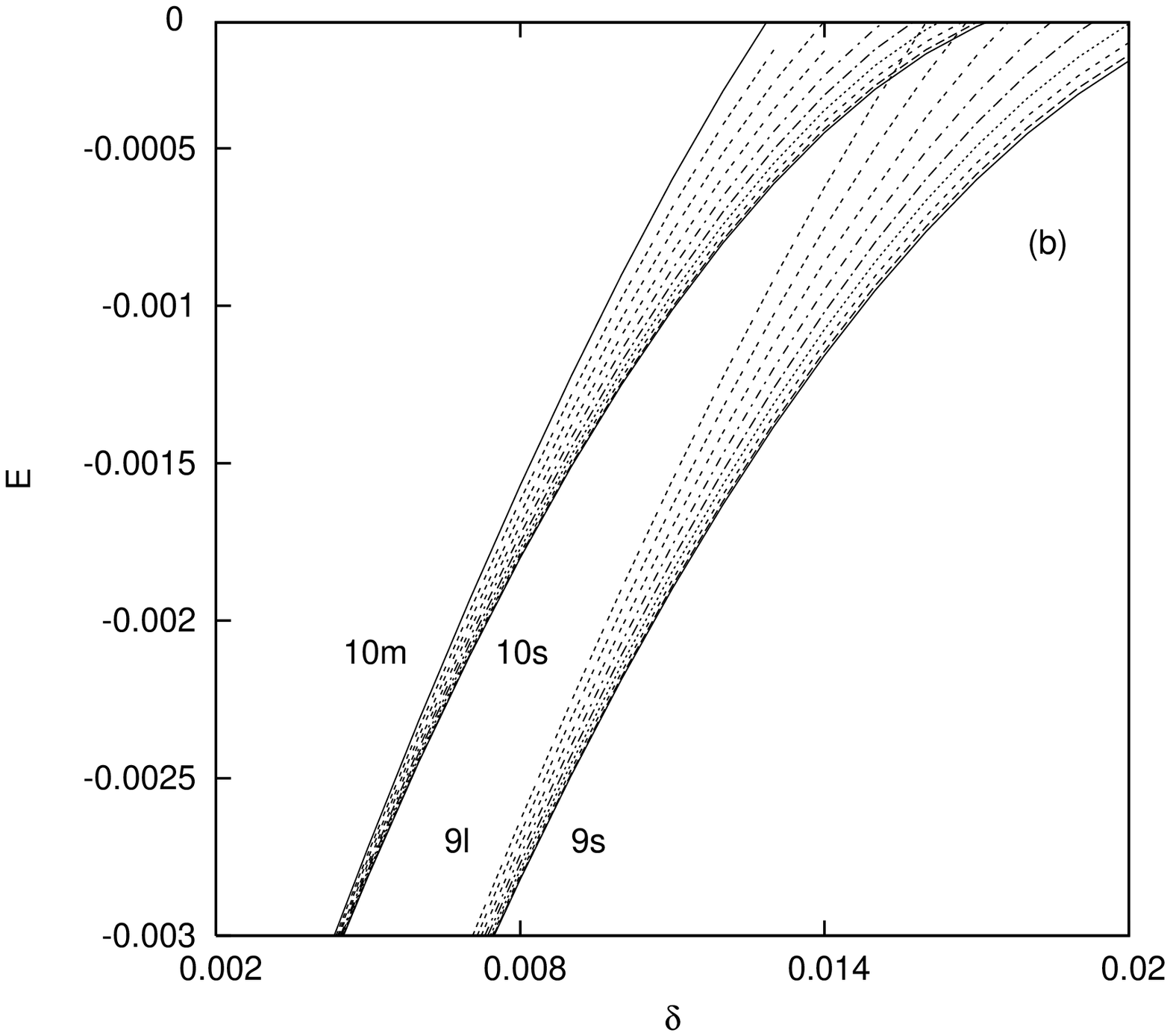}
\end{minipage}%
\caption{Energy eigenvalues (a.u.) of the Hulth\'en potential for (a) $n=7,8$ and 
(b) $n=9,10 $ levels respectively as a function of $\delta$ in the vicinity of 
zero energy.}
\end{figure}

Next in table II we present energies of the representative non-zero angular momentum
states of the Hulth\'en potential with $n=2-6$. As already mentioned, these states do
not offer exact analytical results and a large number of attempts have been made over
the years, e.g., the variational as well as numerical integration [18], strong-coupling
expansion [42], supersymmetric quantum mechanics [29], parameter-free wave function 
approach based on the local properties such as the cusp conditions [31], etc., in 
addition to some of the methods which also dealt with the $\ell=0$ case such as 
[16,19,24,28,31]. Other works include [7,43,44] and the best results are quoted here 
for comparison. The $\delta_c$s in these cases cannot be obtained by a simple form and
an approximate analytical expression was put forth by [42],
\begin{equation}
\delta_c=1/[n\sqrt2 + 0.1645\ell +0.0983\ell/n]^2
\end{equation}
which offered results rather in good agreement with the numerically determined values
[18]. These values from [18] are given in parentheses. Majority of the previous
works have dealt with the weak coupling regions. We have chosen $\delta$ values
for all the states quite wide; from weak to strong. A uniform accuracy is maintained
for all these states in the whole region of interaction, unlike some other previous
calculations which encountered difficulties in the stronger limits (e.g., the break-
down of the shifted 1/N expansion method for ground state with $\delta > 1.2$ [23]). 
In the weaker limit, our computed eigenvalues are superior to all other results except 
the accurate variational calculations [16]. However in this work, we have gone beyond
the interaction region considered in [16] or in any other previous calculation so far
for some of these states ($2p$, $3p$, $4d$, $4f$) and no results could be found for
these states for direct comparison. Now Fig. 1 depicts the variation of energy 
eigenvalues with screening parameter for all the states belonging to $n=7,8$ (left)
and $n=9,10$ (right) respectively. For small values of the principal quantum number, 
there is a resemblance of the energy orderings as those with the Coulomb potentials,
but with an increase in $n$, significant deviations from the Coulomb potential ordering
and complex level crossings observed in the vicinity of zero energy. This is more 
pronounced in the latter case ($9k$, $9l$ mixing heavily with $10s$, $10p$, $10d$ 
and $10f$ at around $\delta=0.015-0.017$) and for higher $n$, there is a gradual 
increase in the probability of energy ordering becoming more complex, which make their
accurate calculations quite difficult. We also notice that for a particular value of 
$n$, the separation between states with different values of $l$ increases with 
$\delta$. Additionally now in table III we give the calculated eigenvalues for all 
the states belonging to $\ell \neq 0, n=8$ and 10 at $\delta=0.02$ 0.01 values 
respectively. While states with relatively higher $n$ values have been studied for 
the Yukawa potential (see later), no results have been reported so far for such higher
states of the Hulth\'en potential. For the sake of completeness, here also the 
$\delta_c$ values are quoted from [18].  

\begingroup
\squeezetable
\begin{table}
\caption {\label{tab:table3} The calculated negative eigenvalues (a.u.) of Hulth\'en 
potential for $\ell \neq 0, n=8,10$ states at $\delta$ values 0.02 and 0.01 
respectively. Numbers in the parentheses denote the $\delta_c$ values taken from [18].} 
\begin{ruledtabular}
\begin{tabular}{llllll}
 State & $\delta$ & $-$Energy & State &  $\delta$ & $-$Energy   \\ \hline 
$8p(0.030)$ & 0.02 & 0.0009868327076 & $10p(0.019)$ & 0.01 & 0.0012427752748 \\
$8d(0.028)$ & 0.02 & 0.0009349530511 & $10d(0.018)$ & 0.01 & 0.0012282621767 \\
$8f(0.026)$ & 0.02 & 0.0008556949061 & $10f(0.017)$ & 0.01 & 0.0012063302045 \\
$8g(0.025)$ & 0.02 & 0.0007470911124 & $10g(0.017)$ & 0.01 & 0.0011767752007 \\
$8h(0.023)$ & 0.02 & 0.0006060126055 & $10h(0.016)$ & 0.01 & 0.0011393080783 \\
$8i(0.022)$ & 0.02 & 0.0004274312523 & $10i(0.015)$ & 0.01 & 0.0010935375130 \\
$8k(0.021)$ & 0.02 & 0.0002027526409 & $10k(0.015)$ & 0.01 & 0.0010389439316 \\
            &      &                 & $10l(0.014)$ & 0.01 & 0.0009748402143 \\
            &      &                 & $10m(0.013)$ & 0.01 & 0.0009003110142 \\  
\end{tabular}
\end{ruledtabular}
\end{table}
\endgroup

\begingroup
\squeezetable
\begin{table}
\caption {\label{tab:table4}The calculated negative energy eigenvalues (a. u.) of
the Yukawa potential as a function of $\lambda$ for representative $n \leq 6$ states 
along with the literature data. Numbers in the parentheses denote $\lambda_c$ values 
quoted from [30].} 
\begin{ruledtabular}
\begin{tabular}{ccllcll}
State & $\lambda$ & \multicolumn{2}{c}{$-$Energy} & $\lambda$ & 
\multicolumn{2}{c}{$-$Energy} \\ 
\cline{3-4} \cline{6-7}
   &  & This work & Literature &  & This work & Literature \\    \hline 
$1s$(1.1906) & 0.01  & 0.40705803061340 & 
0.40705803061340\footnotemark[1]$^,$\footnotemark[2]$^,$\footnotemark[3] & 1.19 
            & 0.00000010303196 & 0.00000010303196\footnotemark[3] \\   
$2p$(0.2202)& 0.01 & 0.11524522409056 & 
                     0.11524522409056\footnotemark[2]$^,$\footnotemark[3] & 0.22
      & 0.00002869724498 & 0.00002869\footnotemark[2],0.000026\footnotemark[3]  \\   
$3p$(0.1127)& 0.01 & 0.04615310482916 & 0.04616\footnotemark[4],0.04615\footnotemark[5]
            & 0.11 & 0.00022634084060 &                      \\
$3d$(0.0913)& 0.01 & 0.04606145416065 & 0.04606\footnotemark[4]$^,$\footnotemark[5]
            & 0.09 & 0.00031291350263 &                      \\
$4d$(0.0581)& 0.01 & 0.02222779248980 & 0.02222779248980\footnotemark[2] 
           & 0.055 & 0.00049188376726 &                      \\
$4f$(0.0498)& 0.005& 0.02646809608410 & 0.02647\footnotemark[4],0.02645\footnotemark[5]
           & 0.045 & 0.00148735974333 &  0.00146\footnotemark[5] \\    
$5f$(0.0354)& 0.01 & 0.01142540016608 & 0.01142540016608\footnotemark[2] & 0.035
            & 0.00006899773341 &                  \\
$5g$(0.0313)& 0.01 & 0.01126616478845 & 0.01126616478845\footnotemark[2] & 0.031
            & 0.00009981963916 &                  \\
$6f$(0.0264)& 0.005 & 0.00944274896286 & 0.00944274896286\footnotemark[2] & 0.025
            & 0.00019933872619 & 0.00020\footnotemark[4]            \\
$6g$(0.0238)& 0.005 & 0.00940059908613 & 0.00940059908613\footnotemark[2] & 0.022
            & 0.00042240106329 &                  \\
$6h$(0.0215)& 0.005 & 0.00934767158207 & 0.00934767158207\footnotemark[2] & 0.021
            & 0.00016299459024 &                  \\
\end{tabular}
\end{ruledtabular}
\footnotetext[1] {Ref. [31].}
\footnotetext[2] {Ref. [16].}
\footnotetext[3] {Ref. [21].}
\footnotetext[4] {Ref. [30].}
\footnotetext[5] {Ref. [33].}
\end{table}
\endgroup

\begin{figure}
\begin{minipage}[c]{0.40\textwidth}
\centering
\includegraphics[scale=0.45]{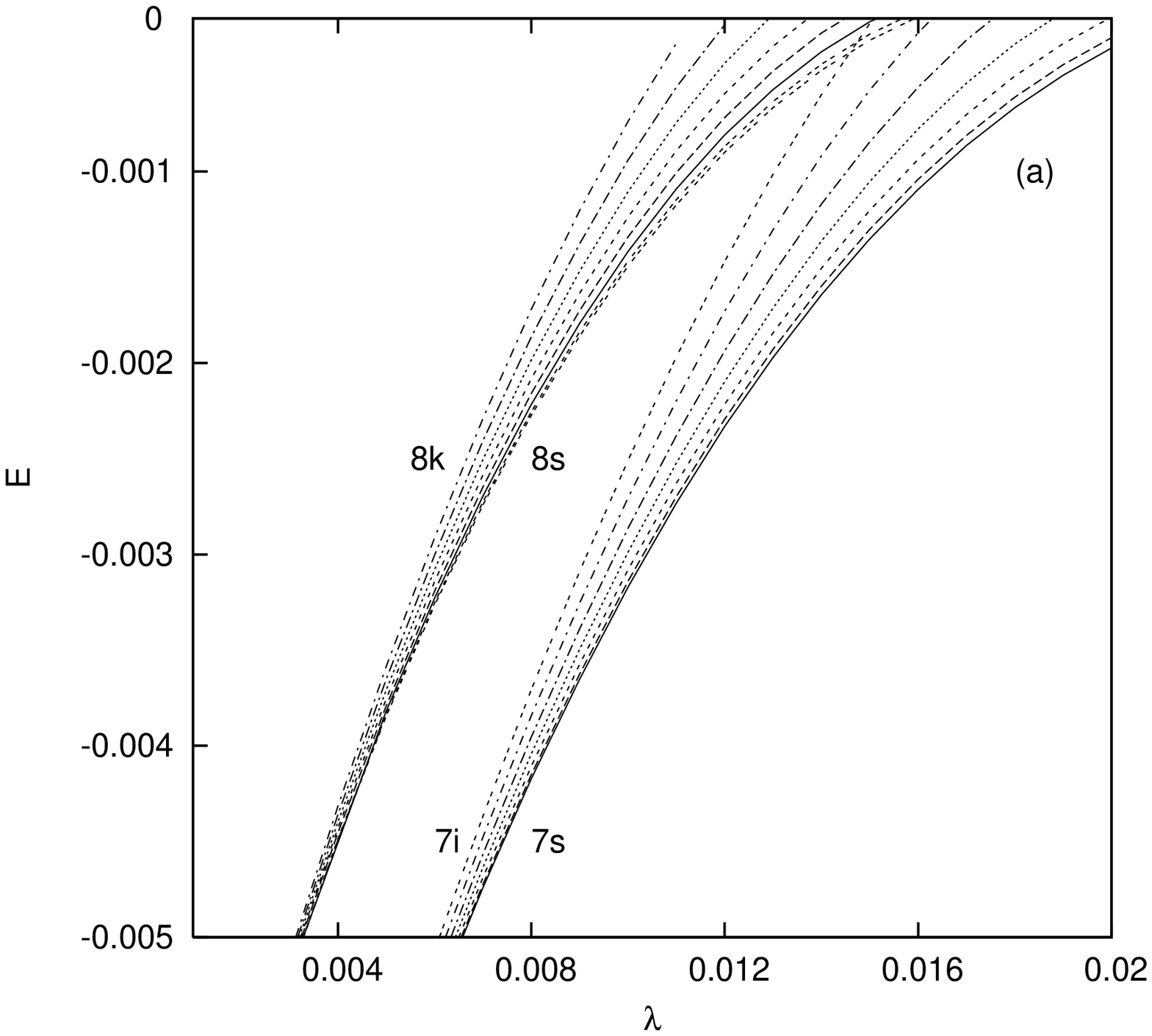}
\end{minipage}%
\hspace{0.5in}
\begin{minipage}[c]{0.40\textwidth}
\centering
\includegraphics[scale=0.45]{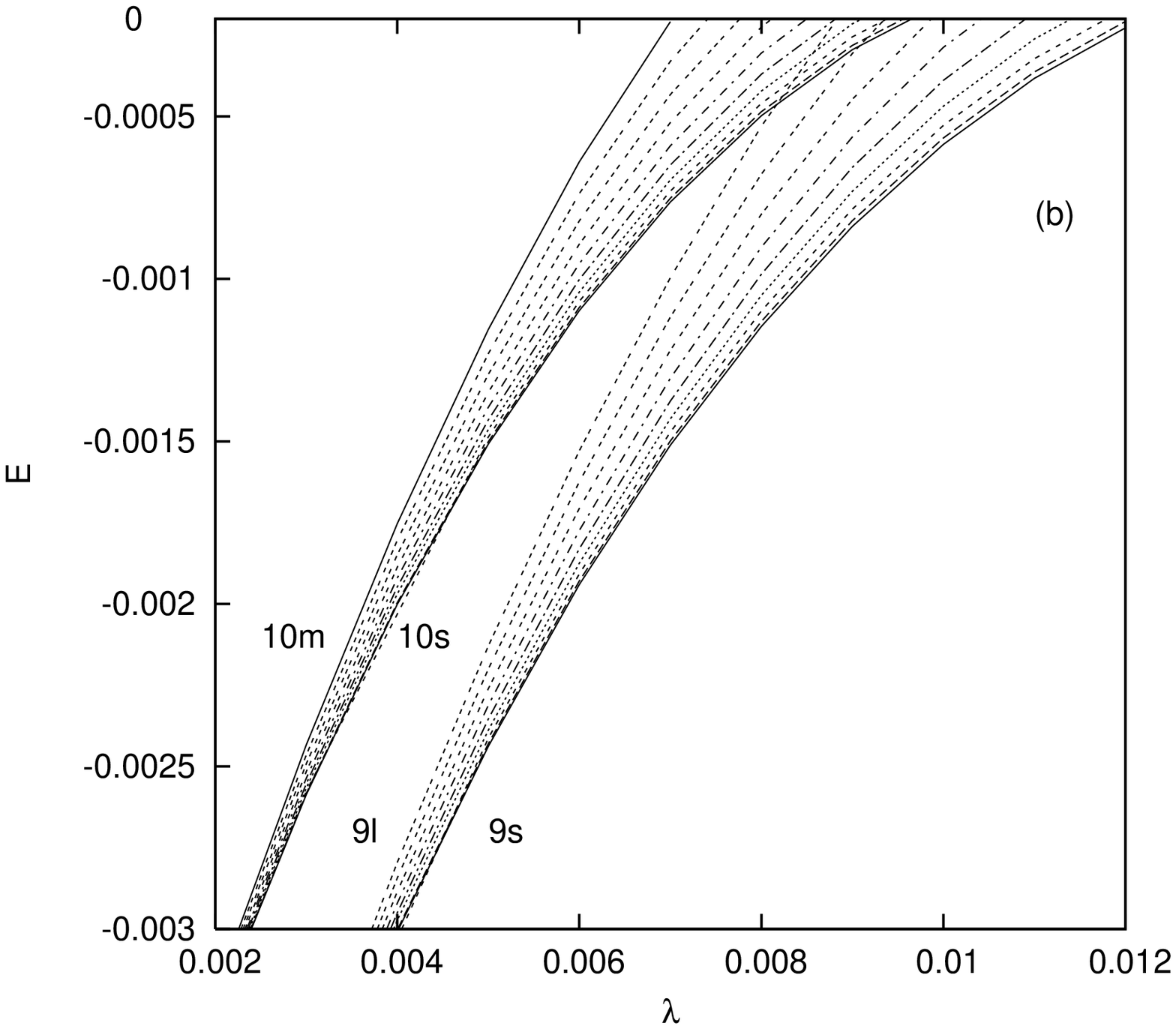}
\end{minipage}%
\caption{Energy eigenvalues (a.u.) of the Yukawa potential for (a) $n=7,8$ and (b) 
$n=9,10$ (right) levels respectively as a function of $\lambda$ in the vicinity of 
zero energy.}
\end{figure}

Now we turn our focus on to the Yukawa potentials. Table IV presents the calculated 
eigenvalues of some representative states with $n \leq 6$ at selected values of the 
screening parameter (weak and strong screenings in the left and right respectively). 
Lower states have been examined by many methods; e.g., Rayleigh-Schr\"odinger 
perturbation expansion [21], variational methods [6,15,16], Pad\'e approximations 
[20], shifted 1/N expansions [22,25], numerical calculations through direct 
integration of the SE [30] 
or by the Ritz method [31], etc. Other works include [12,33]. However, excepting a
few of these (like [21] or [16]), majority of them produce good-quality results in
the weaker regions of interaction, and are often fraught with difficulties in the
stronger regions (e.g., the shifted 1/N expansion [22] runs into trouble for 
$\lambda$ near the critical values). The best available literature data are given 
for comparison. The numerically determined critical screening parameters 
($\lambda_c$)s for these states are quoted from [30]. First of all, very good agreement
is observed for all these states with the best available literature data. Once again
a wide range of interaction region has been considered and the converged results are
uniformly accurate for all of these states for arbitrary values of $\lambda$. We note
that, accurate numerical results [31] have been reported for the $1s$, $2s$ and $2p$
states in the weaker coupling region. Our $1s$ results are as accurate as those of 
[31], while those for $2s$ and $2p$ states are superior to [31]. As in the case of 
Hulth\'en potential, results of [16] are more accurate than ours for smaller $\lambda$, 
but we have obtained better results in the stronger regions (e. g., the $1s, 2p, 3s, 6s,
6p$ states with $\lambda= 1.15, 0.22, 0.12, 0.03, 0.03$ respectively). Very accurate 
energies were reported in [21] for $1s, 2s, 2p$ states having both small and large
$\lambda$s. Results of [21] are better than ours for the first two $s$ states, but
for the $2p$ state present results deviate considerably from [21] in the stronger
region (e. g., $\lambda=0.21, 0.22$) and we believe these are the better results. Some
of these states have not been calculated by any method other than those of [30,33], and 
the GPS results improve those values dramatically. Thus to our knowledge, these appear 
to be the most accurate results for these states (except $1s,2s$) in the regions closest
to the critical domain. In Fig. 2 we graphically show the dependence of the energy 
orderings of all the $n=7,8$ (left) and $n=9,10$ (right) states of the Yukawa potential
on the screening parameter $\lambda$ in the vicinity of E=0. Essentially similar 
qualitative features are observed as in Fig. 1 for the Hulth\'en potential, {\em viz.} 
(a) gradually more complex level crossings as $n$ increases and (b) the energy splitting 
between the states with different values of $\ell$ for a given value of $n$ increases 
with an increase in the screening parameter. Table V gives all the eigenstates for 
$n=9$ and 10 at selected values of $\lambda$ respectively. As $n$ and $\ell$ increase, 
calculation of these states become 
progressively difficult, and only two attempts have been made so far to study the 
$7s-9\ell$ states, {\em viz.,} the direct numerical integration of SE [30] as well as 
the shifted 1/N expansion [22]. The former results are more accurate than the latter and 
these are quoted. While their works [30] estimated these states fairly accurately and 
still by far the most reliable values reported in the literature, clearly the present 
results are much more improved in accuracy. The $\lambda_c$ values in these cases are 
taken from [30]. No attempts are known for any of the states with $n > 9$ and here we 
have given them for the first time which may constitute a useful reference for future 
studies. Also in this table are included the results for some of the representative 
$\ell=0$ states with $n$ up to 17 and no comparisons could be made because of the 
lack of literature data. 

\begingroup
\squeezetable
\begin{table}
\caption {\label{tab:table5}Comparison of the negative eigenvalues (a. u.) of 
Yukawa potential for $n >6$ states at selected values of $\lambda$. Numbers in the
parentheses in column 1 denote $\lambda_c$ values and in all other columns, the 
numerical eigenvalues; both from [30].} 
\begin{ruledtabular}
\begin{tabular}{lcllccll}
State & $\lambda$ & \multicolumn{2}{c}{$-$Energy} & State &  $\lambda$ & 
\multicolumn{2}{c}{$-$Energy}   \\ 
\cline{3-4} \cline{7-8}
   &  & This work & Literature [30] &  &   &   This work & Literature \\    \hline 
$9s$(0.016)  & 0.01  & 0.0005858247612 & 0.000585 & $10s$ & 0.005 & 0.0015083559307 &   \\
$9p$(0.015)  & 0.01  & 0.0005665076261 & 0.000565 & $10p$ & 0.005 & 0.0015009235029 &   \\
$9d$(0.014)  & 0.01  & 0.0005276644203 & 0.00053  & $10d$ & 0.005 & 0.0014860116240 &   \\
$9f$(0.013)  & 0.01  & 0.0004688490636 & 0.00047  & $10f$ & 0.005 & 0.0014635239275 &   \\
$9g$(0.012)  & 0.01  & 0.0003893108558 & 0.00039  & $10g$ & 0.005 & 0.0014333097805 &   \\
$9h$(0.011)  & 0.01  & 0.0002878564558 & 0.00029  & $10h$ & 0.005 & 0.0013951561294 &   \\
$9i$(0.0107) & 0.005 & 0.0022606077422 & 0.00226  & $10i$ & 0.005 & 0.0013487749860 &   \\
$9k$(0.0100) & 0.005 & 0.0021997976659 & 0.00220  & $10k$ & 0.005 & 0.0012937846259 &   \\
$9l$(0.0094) & 0.005 & 0.0021291265596 & 0.00213  & $10l$ & 0.005 & 0.0012296811835 &   \\
$10m$        & 0.005 & 0.0011557947569 &          & $11s$ & 0.002 & 0.002455067336  &   \\
$12s$        & 0.002 & 0.001849081136  &          & $13s$ & 0.002 & 0.001392026936  &   \\
$16s$        & 0.001 & 0.001122878263  &          & $17s$ & 0.001 & 0.000919120394  &   \\
\end{tabular}
\end{ruledtabular}
\end{table}
\endgroup

As a further test of the convergence of eigenfunctions, the calculated density moments
$\langle r^{-1} \rangle$ and $\langle r \rangle$ are given in table VI for a few states
at selected values of the screening parameters of both Hulth\'en (left) and Yukawa 
(right) potentials. The best available numerical results [31] are quoted in the 
parentheses wherever available, and for all these instances, we have obtained 
superior results than the previous reported values in literature. Additionally, the 
left portion of Fig. 3 shows the variation of radial probability distribution functions
for ground states of the Yukawa potential with respect to $\lambda$. Five values of
$\lambda$ are considered, {\em viz.,} 0.1 (low screening), 1.0 and 1.1 (moderate
screening), 1.12 and 1.15 (high screening). It is seen that with an increase in
$\lambda$, the density distribution oozes out to larger values of $r$ and the peak
values are reduced. The right portion of Fig. 3 depicts the density distributions 
for $2s$, $3s$ and $4s$ states of the Yukawa potential with $\lambda=0.01$. As expected
it spreads out to larger $r$ as $n$ increases and the requisite number of nodes are 
present. Analogous features are also observed for the Hulth\'en potential. At this 
stage, a few comments should be made. It has been pointed out [16] that the 
eigenvalues of Coulomb, Hulth\'en and the Yukawa potentials follow the relation,
\begin{equation}
E_n^{\mathrm{coulomb}} \leq E_{n,\ell}^{\mathrm{Hulthen}}(\delta)
\leq E_{n,\ell}^{\mathrm{Yukawa}}(\lambda)
\end{equation}
and this has been verified to be satisfied for all the states considered in this
work. Finally, we mention here that the GPS method employed here possesses the 
simplicity of FD or FE methods and at the same time retain the attractive
features of the basis-set variational methods, such as high accuracy and fast
convergence. It is also known that it guarantees an ``exponential'' (also called 
infinite-order) convergence for a given problem with smooth (infinitely 
differentiable) solutions (which is usually the case) as long as the orthogonal
functions employed belong to a common singular Sturm-Liouville class. Furthermore,
a pseudospectral method with N+1 or N+2 grid points is usually equivalent in accuracy
to the corresponding basis-set expansion method with N basis functions (for a 
detailed account of these and other features of GPS method, see [34-39, 44,45]
and the references therein).

\begingroup
\squeezetable
\begin{table}
\caption {\label{tab:table6} The calculated expectation values of Hulth\'en (left)
and Yukawa (right) potentials for several $s$ states as a function of the screening
parameters. Numbers in the parentheses are taken from [31].} 
\begin{ruledtabular}
\begin{tabular}{c|cll|cll}
State & $\delta$  & $\langle r^{-1} \rangle$ & $\langle r \rangle$ 
      & $\lambda$ & $\langle r^{-1} \rangle$ & $\langle r \rangle$ \\  \hline
$1s$  & 0.1   & 0.998748957029 & 1.502506265664 & 0.5   & 0.867533084978 & 1.806554897095 \\
      &       &                &                &       &(0.867533084)   &(1.806554897)   \\
$2s$  & 0.1   & 0.244953137615 & 6.113636363636 & 0.1   & 0.227996338490 & 6.529952268703 \\
      &       &                &                &       &(0.2279963389)  &(6.52995228)    \\
$3s$  & 0.001 & 0.111109986106 & 13.50011475165 & 0.001 & 0.111104429011 & 13.50068154934 \\  
$16s$ & 0.001 & 0.00387414697  & 386.4792127    & 0.001 & 0.00375246427  & 396.3082404    \\  
$17s$ & 0.001 & 0.00342393411  & 437.0760155    & 0.001 & 0.00329034570  & 450.9901509    \\  
\end{tabular}
\end{ruledtabular}
\end{table}

\section{conclusion}
A detailed study has been made on the accurate eigenvalues, density moments and radial
densities of Hulth\'en and Yukawa potentials by employing the GPS formalism. The 
methodology is simple, efficient, accurate and reliable. Special attention has been paid
to the higher excited states as well the stronger screening effects. All the 55 states
belonging to $n \leq 10$ have been computed with good accuracy and the results are 
compared wherever possible. In the weak coupling regions, our results are comparable 
to all other accurate literature results available (except [21] and [16]), while in the
strong coupling region, present results are noticeably superior to all other existing
results for all states of both these systems excepting the $1s$ and $2s$ states of 
Yukawa potential. The $n > 6$ states of Yukawa potentials are significantly
improved from the best available data available so far, while the same for Hulth\'en
potential are reported here for the first time. In view of the simplicity and accuracy
offered by this method for both these physical systems studied in this work, it is hoped
that this may be equally successful and useful for other singular potentials in various
branches of quantum mechanics. 

\endgroup
\begin{figure}
\begin{minipage}[c]{0.40\textwidth}
\centering
\includegraphics[scale=0.45]{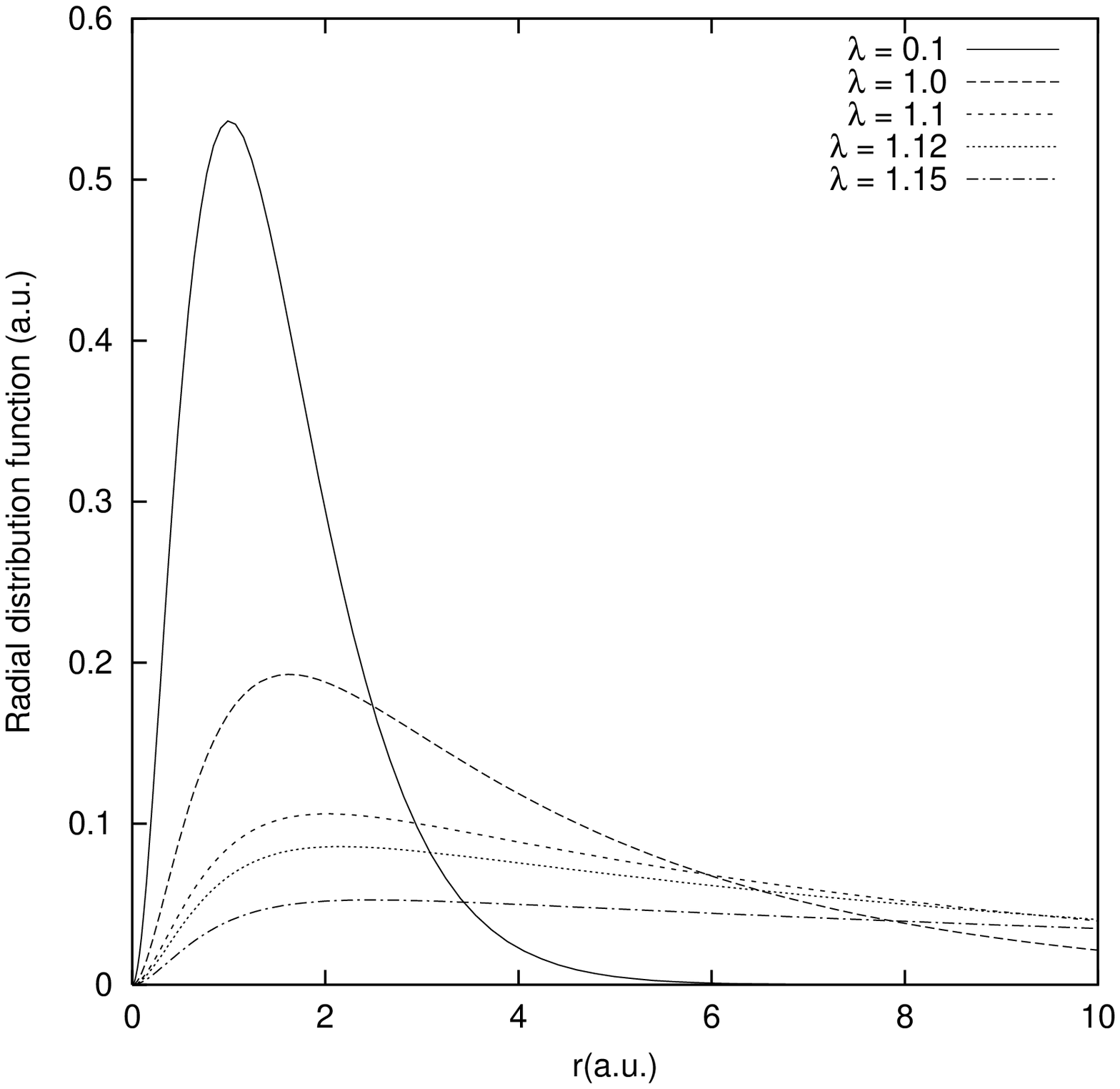}
\end{minipage}%
\hspace{0.5in}
\begin{minipage}[c]{0.40\textwidth}
\centering
\includegraphics[scale=0.45]{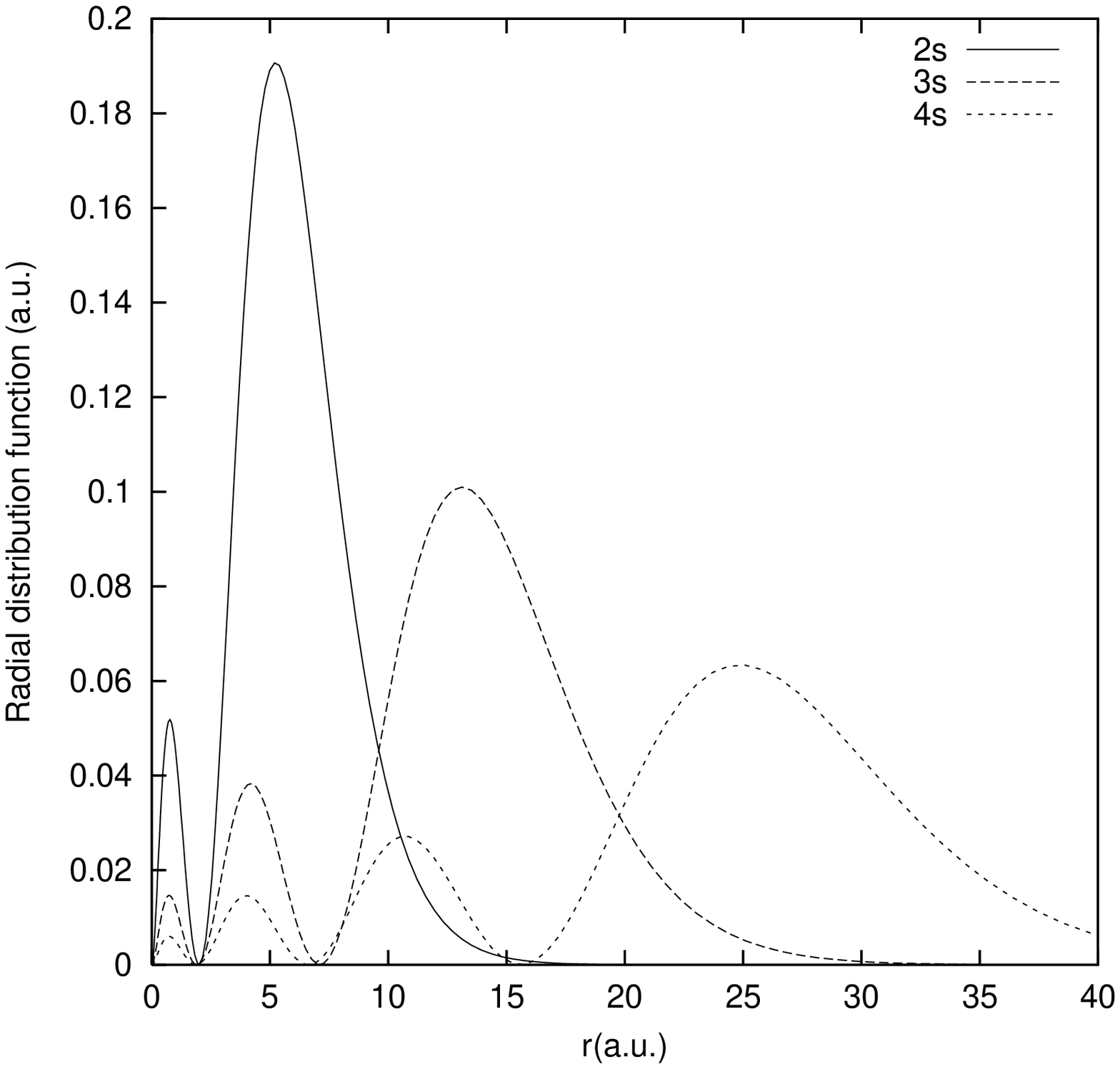}
\end{minipage}%
\caption{The radial densities (a.u.) of Yukawa potential for the ground states with
$\lambda=0.1, 1.0, 1.1, 1.12, 1.15$ (left) and $2s$, $3s$, $4s$ states having 
$\lambda=0.01$ (right) respectively.}
\end{figure}

\begin{acknowledgments}
I gratefully acknowledge the warm hospitality provided by the University of New Brunswick,
Fredericton, NB, Canada. I am greatly thankful to the anonymous referee for numerous 
valuable comments.
\end{acknowledgments}

\end{document}